\documentclass[12pt]{amsart}
\usepackage{geometry}                
\usepackage{graphicx}
\usepackage{amssymb}
\usepackage{epstopdf}
\usepackage{breqn}
\title{Mathematical Issues in Eternal Inflation}
\author{Ikjyot Singh Kohli}
\author{Michael C. Haslam}
\address{Department of Physics and Astronomy, York University, Toronto, Ontario}
\address{Department of Mathematics and Statistics, York University, Toronto, Ontario}
\email{isk@yorku.ca}
\email{mchaslam@mathstat.yorku.ca}
\date{August 30, 2014}                                           

\begin{document}


\begin{abstract}
In this paper, we consider the problem of existence and uniqueness of solutions to the Einstein field equations for a spatially flat FLRW universe in the context of stochastic eternal inflation where the stochastic mechanism is modelled by adding a stochastic forcing term representing Gaussian white noise to the Klein-Gordon equation. We show that under these considerations, the Klein-Gordon equation actually becomes a stochastic differential equation. Therefore, the existence and uniqueness of solutions to Einstein's equations depend on whether the coefficients of this stochastic differential equation obey Lipschitz continuity conditions.  We show that for any choice of $V(\phi)$, the Einstein field equations are not globally well-posed, hence, any solution found to these equations is not guaranteed to be unique. Instead, the coefficients are at best locally Lipschitz continuous in the physical state space of the dynamical variables, which only exist up to a finite explosion time. We further perform Feller's explosion test for an arbitrary power-law inflaton potential and prove that all solutions to the EFE explode in a finite time with probability one. This implies that the mechanism of stochastic inflation thus considered cannot be described to be eternal, since the very concept of eternal inflation implies that the process continues indefinitely. We therefore argue that stochastic inflation based on a stochastic forcing term would not produce an infinite number of universes in some multiverse ensemble. In general, since the Einstein field equations in both situations are not well-posed, we further conclude that the existence of a multiverse via the stochastic eternal inflation mechanism considered in this paper is still very much an open question that will require much deeper investigation.
\end{abstract}

\maketitle

\section{Introduction}
In this paper, we consider the problem of existence and uniqueness of solutions to the Einstein field equations for a spatially flat Friedmann-Lema\^{\i}tre-Robertson-Walker (FLRW) universe in the context of stochastic eternal inflation. Stochastic eternal inflation has received a considerable amount of attention over the past few years as it is one of main motivations behind the multiverse \cite{elliscosmo}. In particular, Miji\'{c} \cite{mijic1} analyzed the boundary conditions of the dynamics of inflation as a relaxation random process and gave a simple proof for the existence of eternal inflation. Salopek and Bond \cite{salobond} studied nonlinear effects of the metric and scalar fields in the context of stochastic inflation. Linde, Linde, and Mezhlumian \cite{lindelindemezh} considered chaotic inflation in theories with effective potentials that behave either as $\phi^n$ or as $e^{\alpha \phi}$. They also performed computer simulations of stochastic processes in the inflationary universe. Linde and Linde  \cite{lindelinde} investigated the global structure of an inflationary universe both by analytical methods and by computer simulations of stochastic processes in the early universe. Susperregi and Mazumdar \cite{susmazum} considered an exponential inflation potential and showed that this theory predicts a uniform distribution for the Planck mass at the end of inflation, for the entire ensemble of universes that undergo stochastic inflation. Vanchurin, Vilenkin, and Winitzki \cite{vanvilwin} investigated methods of inflationary cosmology based on the Fokker-Planck equation of stochastic inflation and direct simulation of inflationary spacetime. Winitzki \cite{winitzki} explored the fractal geometry of spacetime that results from stochastic eternal inflation. Kunze \cite{kunze} considered chaotic inflation on the brane in the context of stochastic inflation. Winitzki \cite{winitzki2} described some issues regarding the time-parameterization dependence in stochastic descriptions of eternal inflation. Gratton and Turok \cite{grattonturok} investigated a simple model of $\lambda \phi^4$ inflation with the goal of analyzing the continuous revitalization of the inflationary process in some regions. Li and Wang \cite{liwang1} used a stochastic approach to investigate a measure for slow-roll eternal inflation. Tom\`{a}s G\`{a}lvez Ghersi, Geshnizjani, Piazza, and Shandera \cite{tomas} used stochastic eternal inflation to analyze the connection between the Einstein equations and the thermodynamic equations. Qiu and Saridakis \cite{qiusarid} used stochastic quantum fluctuations through a phenomenological, Langevin analysis studying whether they can affect entropic inflation eternality. Harlow, Shenker, Stanford, and Susskind \cite{harlowshenker} described a discrete stochastic model of eternal inflation that shares many of the most important features of the corresponding continuum theory. Vanchurin \cite{vanchurin1} developed a dynamical systems approach to model inflation dynamics. Feng, Li, and Saridakis \cite{fenglisaridakis} investigated conditions under which phantom inflation is prevented from being eternal. 

In this paper, we consider the effects of adding a stochastic forcing term in the form of Gaussian white noise to the right-hand-side of the Klein-Gordon equation as is done by Gratton and Turok \cite{grattonturok} in their study of employing the Langevin formalism to eternal inflation.  We show that because this stochastic term is not differentiable at any point, the existence and uniqueness of Einstein's equations depends on the Lipschitz continuity of the coefficients of the corresponding stochastic differential equation. Indeed, in \cite{grattonturok}, the authors also conclude, although through different means, namely that the Langevin approach leads to a monotonically decreasing Hubble constant, which implies that one cannot have eternal inflation in this scenario. Instead, as an alternative to the eternal inflation picture, they propose a model with a constraint on the initial and final states that lead to a picture similar to what is typically associated with eternal inflation.

Throughout, we use geometrized units, such that $G = c = 1$, and as a result of this, all physical quantities have dimensions of powers of length \cite{waldbook}. 

\section{Description of the Model}
We consider a spatially homogeneous and isotropic universe described by the FLRW metric \cite{elliscosmo} as
\begin{equation}
\label{FLRW1}
ds^2 = -dt^2 + a^2(t) \left[dr^2 + f^2(r) \left(d\theta^2 + \sin^2 \theta d \phi^2\right)\right], \quad u^\mu = \delta^{\mu}_{0},
\end{equation}
where 
\begin{equation}
f(r) = (\sinh r, r, sin r) \mbox{ for } K = (-1, 0, +1),
\end{equation}
where $K$ denotes the sign of the curvature of the particular FLRW model under consideration. Namely, $K = -1$ refers to hyperbolic FLRW models, $K = 0$ refers to flat FLRW models, and $K=+1$ refers to positively curved FLRW models.

We assume that energy-momentum tensor is that of a scalar field and has the form \cite{elliscosmo}
\begin{equation}
\label{emtensor}
T_{\phi}^{ab} = \nabla^{a} \phi \nabla^{b} \phi - \left[\frac{1}{2} \nabla_{c} \phi \nabla^{c} \phi + V(\phi)\right]g^{ab}.
\end{equation}

The Einstein field equations that also describe the dynamics of the model are given by the Raychaudhuri and Friedmann equations \cite{elliscosmo},
\begin{eqnarray}
\label{raych}
\dot{H} &=& -H^2 + \frac{1}{3} \left[V(\phi) - \dot{\phi}^2\right],  \\
\label{fried1}
3H^2 &=& V(\phi) + \frac{\dot{\phi}^2}{2} + \frac{1}{2} ^{(3)}R,
\end{eqnarray}
where $H$ is the Hubble parameter, and $^{(3)}R$ is the three-dimensional Ricci scalar, which is constant for the FLRW models.

The contracted Bianchi identities give an evolution equation for the scalar field, $\phi$, which in this case is precisely the Klein-Gordon equation \cite{elliscosmo},
\begin{equation}
\label{kgordon}
\ddot{\phi} + 3H\dot{\phi} + V'(\phi) = 0.
\end{equation}

Together Eqs. \eqref{raych}, \eqref{fried1}, and \eqref{kgordon} fully describe the dynamics of the cosmological model described by Eqs. \eqref{FLRW1} and \eqref{emtensor}.

\section{A Model of Stochastic Eternal Inflation}
Following \cite{elliscosmo}, we note that the classical dynamics of the inflaton dictates that the inflaton always rolls down its potential. However, quantum fluctuations can also drive the inflaton uphill, which causes inflation to last longer in some regions, which, in turn, enlarges the volume of the region. In some regions, the inflaton will remain high enough up the potential hill to maintain acceleration. This is a stochastic scenario that is typically described as \emph{eternal} inflation which is one of the main motivating ideas behind the multiverse concept. We will model this stochastic behaviour following \cite{grattonturok} and \cite{lindelindemezh} where a stochastic forcing term representing Gaussian white noise is added to the right-hand-side of Eq. \eqref{kgordon}. 

As discussed in \cite{grattonturok}, the purpose of this stochastic term is to describe the scalar field fluctuations. Further, this stochastic term arises because of initially small-scale quantum fluctuations across the de Sitter horizon. It is typically claimed in inflationary studies that the physical size of the region that is known to be inflating increases quite rapidly. However, this statement only applies when the initial region is so large that causal influences coming from outside the initial region cannot propagate far enough into it to stop inflationary process. Therefore, the stochastic model under consideration only applies to a region that is small enough to lie within the past light cone of a future observer, and large enough to remain in an inflating state well into the future. The reader is encouraged to see Figure 1 in \cite{grattonturok}. 

We will denoting the stochastic forcing term by $H^{5/2}\eta(t)$ \cite{grattonturok}, we find that the dynamical equations \eqref{raych} and \eqref{kgordon} can be written as coupled first-order ordinary differential equations in the form:
\begin{eqnarray}
\label{phidot}
\frac{d \phi}{dt} &=& f, \\
\label{Hdot}
\frac{d H}{dt} &=& -H^2 + \frac{1}{3} \left[V(\phi) -f^2\right], \\
\label{fdot}
\frac{df}{dt} &=& H^{5/2} \eta(t) - 3Hf - V'(\phi), \\
\label{constr1}
3H^2 &=& V(\phi) + \frac{f^2}{2} + \frac{1}{2} ^{(3)}R,
\end{eqnarray}
where Eq. \eqref{constr1} is the generalized Friedmann equation and acts as a constraint on the initial conditions.
Analyzing the system \eqref{phidot}-\eqref{fdot}, we see that the right hand side of Eq. \eqref{fdot} is not $C^{1}$ because the stochastic function $\eta(t)$ is not anywhere differentiable. To make this point more precise, we describe some properties of the stochastic function $\eta(t)$. Following \cite{Longtin:2010},we note that  $\eta(t)$ is defined as the time derivative of the Wiener process, which we denote by $W$, that is,
\begin{equation}
\label{wiendef}
\frac{dW}{dt} = \eta(t),
\end{equation}
where $W(0) = 0$ by definition. As noted in \cite{grattonturok}, $\eta(t)$ also satisfies the autocorrelation relation
\begin{equation}
\langle \eta(t) \eta(t') \rangle = \delta(t - t').
\end{equation}
One has to be careful with the definition of $\eta(t)$ as given in Eq. \eqref{wiendef} since $W$ is nowhere differentiable. In fact, for completeness, following \cite{wiersema} we now state some properties of $W$ based on the so-called L\'{e}vy characterization:
\begin{enumerate}
\item The path of $W$ is continuous and starts at 0,
\item $W$ is a martingale and $[dW(t)]^2 = dt$,
\item The increment of $W$ over time period $[s,t]$ is normally distributed, with mean $0$ and variance $(t-s)$,
\item The increments of $W$ over non-overlapping time periods are independent.
\end{enumerate}
Note that, we will not go into extensive detail about martingales. The interested reader is asked to consult \cite{wiersema} for more details on martingales and their properties. Based on the arguments provided in \cite{wiersema}, we will now show that $W$ is not anywhere differentiable, that is, not $C^{1}$ anywhere. Consider a time interval of length $\Delta t = 1/n$ starting at $t$. We will define the rate of change over an infinitesimal time interval $[t, t + \Delta t]$ is
\begin{equation}
X_{n} \equiv \frac{W(t + \Delta t) - W(t)}{\Delta t} = \frac{W(t + \frac{1}{n} - W(t)}{\frac{1}{n}} = n \left[W \left(t + \frac{1}{n}\right) - W(t)\right].
\end{equation}
Therefore, $X_{n}$ is normally distributed with expectation value $0$, variance $n$, and of course, standard deviation, $\sqrt{n}$. It is clear then that $X_{n}$ has the same probability distribution as $\sqrt{n}Z$, where $Z$ is the standard normal distribution. To analyze the $C^{1}$ properties, we must see what happens to $X_{n}$ as $\Delta t \to 0$, in other words as $n \to \infty$. For any $k > 0$, let $X_{n} = \sqrt{n} Z$, then
\begin{equation}
\mathbb{P} \left[ \left\| X_{n} \right\| > k\right] = \mathbb{P} \left[  \left\| Z \right\| > \frac{k}{\sqrt{n}}\right].
\end{equation}
Clearly, as $n \to \infty$, $k / \sqrt{n} \to 0$, so we have that
\begin{equation}
 \mathbb{P} \left[  \left\| Z \right\| > \frac{k}{\sqrt{n}}\right] \to \mathbb{P} \left[ \left\| Z \right\| > 0 \right] = 1.
\end{equation}
The point is that we can choose $k$ to be arbitrarily large, so that the rate of change at time $t$ is not finite, and therefore, $W$ is not differentiable at $t$. Since $t$ is arbitrary, one concludes that $W$ is \emph{nowhere differentiable}. 

\section{Dynamical Equations}
One notices that the Einstein field equations in the present context yield a coupled system of nonlinear ordinary differential equations \eqref{phidot}-\eqref{fdot}. As stated from the onset, our goal in this paper is to analyze the notion of existence and uniqueness of solutions to these equations and to determine the existence of the possibility of exploding solutions within the context of the powerful theorems related to stochastic differential equations. The issue, however, is that all of the theorems concerning the aforementioned properties of such equations that exist in the present literature only deal with one-dimensional stochastic processes. Clearly, Eq. \eqref{fdot} is not a one-dimensional process, as its drift term is coupled to the other differential equations of the system. The interested reader is encouraged to refer to \cite{karatzas} and references therein for a careful review of these ideas. We therefore will employ the powerful technique of re-writing our system of equations in terms of expansion-normalized variables \cite{ellis} that will effectively lead to us obtaining a single stochastic differential equation that describes the dynamics of the system. 

We define expansion-normalized variables denoted by $X,Y$ as
\begin{equation}
\label{eq:Xdef}
X = \sqrt{\frac{V}{3}}\frac{1}{H}, \quad Y = \frac{\dot{\phi}}{\sqrt{6} H}.
\end{equation}
As is done in \cite{ellis}, we additionally introduce a dimensionless time variable $\tau$, such that
\begin{equation}
\label{eq:dimtime}
\frac{dt}{d\tau} = \frac{1}{H}.
\end{equation}

Substituting Eqs. \eqref{eq:Xdef} and \eqref{eq:dimtime} into Eqs. \eqref{phidot}-\eqref{constr1}, one obtains the following dynamical system
\begin{eqnarray}
\label{eq:Xev}
X' &=& X \left(1 + q + Y \lambda\right), \\
\label{eq:Yev}
Y' &=& \mathcal{N} - 3 Y - \frac{1}{3} \left(1-Y^2\right) + Y\left(1+q\right),
\end{eqnarray}
subject to the constraint
\begin{equation}
\label{eq:constr2}
X^2 + Y^2 = 1.
\end{equation}
Note that $q$ is the deceleration parameter found through Raychaudhuri's equation and is in this case
\begin{equation}
\label{eq:qdef}
q = 2Y^2 - X^2.
\end{equation}
In addition, we have also defined an expansion-normalized Gaussian noise term
\begin{equation}
\mathcal{N}(\tau) = \frac{\eta}{3\sqrt{6H}}.
\end{equation}
where $\mathcal{N}$ is the formal time derivative of the Wiener process $W(\tau)$ as explained above. One sees that the motivation for this definition of $\mathcal{N}$ is in fact purely physical. As discussed in \cite{grattonturok}, the dimensions of $\eta(t)$ are length to the power one half (in our geometrized units). We scale $\mathcal{N}$ according to this notion so that it is a dimensionless variable.

Following \cite{wagstaff}, we also have defined a model-dependent dimensionless parameter,
\begin{equation}
\label{eq:lambda}
\lambda = \sqrt{\frac{3}{2}} \frac{V'}{V}.
\end{equation}
Note that in the definition of $\lambda$ in Eq. \eqref{eq:lambda}, we have set the Planck mass to unity as per our choice of units. 

Making use of Eqs. \eqref{eq:constr2} and \eqref{eq:qdef} in Eqs. \eqref{eq:Xev} and \eqref{eq:Yev}, we see that the equations decouple,
\begin{eqnarray}
\label{eq:Xev1}
X' &=& X \left(3 - 3X^2 + \sqrt{1 - X^2}\lambda\right), \\
\label{eq:Yev1}
Y' &=& \left(Y^2-1\right)\left(3 Y  + \lambda\right) + \mathcal{N}(\tau).
\end{eqnarray}
One sees then, that the stochastic dynamics are entirely represented by Eq. \eqref{eq:Yev1}.

We will return to analyzing Eq. \eqref{eq:Xev1} later in the paper. For now, we will focus our attention on Eq. \eqref{eq:Yev1}. In doing so, it is important to note that the state space corresponding to Eq. \eqref{eq:Yev1} is not $Y \in \mathbb{R}$. Notice that from Eq. \eqref{emtensor}, that the energy-momentum tensor is that of a perfect fluid with energy density
\begin{equation}
\label{eq:mudef}
\mu = \frac{1}{2}\dot{\phi}^2 + V(\phi),
\end{equation}
and pressure
\begin{equation}
\label{eq:pdef}
p = \frac{1}{2}\dot{\phi}^2 - V(\phi).
\end{equation}
Following \cite{elliscosmo}, we note that for a non-negative potential energy $V \geq 0$, one has the restriction that 
\begin{equation}
\label{eq:restr1}
\frac{p}{\mu} \leq 1.
\end{equation}
Substituting Eqs. \eqref{eq:mudef}, \eqref{eq:pdef}, \eqref{eq:Xdef}, \eqref{eq:constr2}, and \eqref{eq:qdef} into Eq. \eqref{eq:restr1}, we obtain that
\begin{equation}
\label{eq:statespace1}
-1 \leq Y \leq 1,
\end{equation}
which defines the state space for Eq. \eqref{eq:Yev1}.

\section{Existence and Uniqueness of Solutions}
In this section, we analyze in detail the solutions to the stochastic differential equation \eqref{eq:Yev1}, with the state space defined in \eqref{eq:statespace1}. Using Eq. \eqref{wiendef}, we write stochastic differential equation \eqref{eq:Yev1} in Ito form as
\begin{equation}
\label{eq:Yev2}
dY = \left[ \left(Y^2-1\right)\left(3Y + \lambda\right)\right] d\tau + dW,
\end{equation}
where $-1 \leq Y \leq 1$.

We wish to analyze with respect to this equation whether there exists solutions in the domain \eqref{eq:statespace1}, whether they are unique, and whether there exists some finite time where solutions explode, that is, rapidly converge to infinity with unitary probability. To answer these questions, we state some essential theorems and definitions from stochastic differential equations theory.  

We first note that the coefficient of $d\tau$ in Eq. \eqref{eq:Yev2} is known as the drift coefficient, which we will denote by $b(\tau,Y_{\tau})$. Further, the coefficient of $dW$ is known as the dispersion coefficient, which we will denote by $\sigma(\tau, Y_{\tau})$. It is clear from Eq. \eqref{eq:Yev2} that
\begin{equation}
\label{eq:sdecoeffs}
b\left(\tau, Y_{\tau}\right) = \left(Y^2-1\right)\left(3Y + \lambda\right), \quad \sigma\left(\tau, Y_{\tau}\right) = 1, \quad -1 \leq Y \leq 1.
\end{equation}

Following \cite{marsdencalc}, we note that a function $f$ is \emph{locally} Lipschitz continuous if and only if for $f: A \subset \mathbb{R}^{n} \to \mathbb{R}^{m}$ satisfies $\left\|f(\mathbf{x}) - f(\mathbf{y})\right\| \leq K \left\|\mathbf{x} - \mathbf{y}\right\|$ for all $\mathbf{x}, \mathbf{y} \in A$ for $K > 0$. 

Similarly, a function $f$ is \emph{globally} Lipschitz continuous if one takes according to the previous definition $A = \mathbb{R}^{n}$, that is, the domain is the whole space itself. 

With these definitions in mind, we note that if the coefficients $\mathbf{b}$ and $\mathbf{\sigma}$ as defined in Eq. \eqref{eq:sdecoeffs} are locally Lipschitz continuous then one can only guarantee existence up to an explosion time and uniqueness of solutions in a weak sense. Further, by Feller's explosion test (discussed below) the solutions to a stochastic differential equation with only locally Lipschitz continuous functions as coefficients may explode in a finite time with probability one. We show below that indeed, by Feller's explosion test, that solutions to the field equations within the domain \eqref{eq:statespace1} do indeed explode within a finite time with unitary probability.

Let us look at the function $b(\tau, Y_{\tau})$ from Eq. \eqref{eq:sdecoeffs}. Clearly from the above definition of Lipschitz continuity, one has that for any domain $A \subseteq Y \in \left[-1,1\right]$, there clearly exists a $K > 0$ such  that
\begin{equation}
\left\| \left(x^2-1\right)\left(3x+\lambda\right) - \left(y^2-1\right)\left(3y+\lambda\right) \right\| \leq K \left\| x - y \right\|, \quad x ,y \in A \subseteq Y \in \left[-1,1\right].
\end{equation}
In addition, note that indeed the interval $\left[-1,1\right]$ is closed, so the derivative of $b$ in Eq. \eqref{eq:sdecoeffs}, will be bounded on this interval as well, and by definition, is therefore, locally Lipschitz continuous. 

However, it is quite clear, that if one extends the domain such that $A = \mathbb{R}$, then there exists no $K > 0$ such that
\begin{equation}
\left\| \left(x^2-1\right)\left(3x+\lambda\right) - \left(y^2-1\right)\left(3y+\lambda\right) \right\| \leq K \left\| x - y \right\|, \quad x ,y \in A = \mathbb{R},
\end{equation}
since the polynomial function $b$ is unbounded on $\mathbb{R}$ and one can arbitrarily make the left-hand-side of this inequality large.

Therefore, the coefficients of the stochastic differential equation \eqref{eq:Yev2} are not both globally Lipschitz continuous, and one therefore does not have global existence or uniqueness of solutions to the field equations. The coefficients are however, both locally Lipschitz continuous over the state space $Y \in \left[-1,1\right]$ which means that one at most can have a solution up to an explosion time. In the section that follows, we will show using Feller's explosion test that the solutions to the field equations in this context indeed explode in a finite time over the state space $Y \in \left[-1,1\right]$ with unitary probability. 

From \cite{karatzas}, one defines the sequence
\begin{equation}
S_{n} = \inf \left\{ \tau \geq 0 ; \left\|Y_{\tau}\right\| \geq n\right\}.
\end{equation}
The explosion time for $Y$ is then defined as
\begin{equation}
S = \lim_{n \to \infty} S_{n}.
\end{equation}
We now define the functions $p(x)$, $v(x)$ such that
\begin{equation}
\label{eq:px}
p(x) = \int_{\zeta}^{x} \exp \left[-2 \int_{\zeta}^{s} \frac{ b(r) dr}{\sigma^2(r)}\right]ds, 
\end{equation}
\begin{equation}
\label{eq:vx}
v(x) = \int_{\zeta}^{x} p'(y) \int_{\zeta}^{y} \frac{2 dz}{p'(z) \sigma^2(z)} dy,
\end{equation}
where $\zeta \in (-1,1)$, that is, $\zeta$ assumes a value over the physical state space of the problem as discussed earlier. Feller's test \cite{karatzas} \cite{leonja} then says that suppose that for $b, \sigma: (-1,1) \to \mathbb{R}$, are continuous functions and $\sigma^2 > 0 \in (-1,1)$. The explosion time $\tau$ of the solution $Y$ of Eq. \eqref{eq:Yev2} is finite with probability 1 if and only if any one of the following conditions hold:
\begin{eqnarray}
\label{eq:cond1}
\lim_{x \to 1} v(x) < \infty, &\quad& \lim_{x \to -1} v(x) < \infty, \\
\label{eq:cond2}
\lim_{x \to 1} v(x) < \infty, &\quad& \lim_{x \to -1} p(x) = -\infty, \\
\label{eq:cond3}
\lim_{x \to -1} v(x) < \infty, &\quad& \lim_{x \to 1} p(x) = \infty.
\end{eqnarray}
Per the above definition of the explosion time $S$, the condition for an explosion in finite time is defined probabilistically, 
\begin{equation}
P[S < \infty] = 1.
\end{equation}

Substituting the expressions for $b$ and $\sigma$ from Eq. \eqref{eq:sdecoeffs} into Eqs. \eqref{eq:px} and \eqref{eq:vx}, one obtains
\begin{equation}
\label{eq:newpx}
p(x) = \int_{\zeta }^x \exp \left[-2 \left(-\frac{3 \zeta ^4}{4}+\frac{\zeta ^3 \lambda }{3}+\frac{3 \zeta ^2}{2}-\zeta  \lambda +\frac{3 s^4}{4}-\frac{\lambda  s^3}{3}-\frac{3 s^2}{2}+\lambda  s\right)\right] \, ds,
\end{equation}
where $\zeta \in (-1,1)$ and from which it follows that
\begin{equation}
\label{eq:ppx}
p'(x) = \exp \left[-2 \left(-\frac{3 \zeta ^4}{4}+\frac{\zeta ^3 \lambda }{3}+\frac{3 \zeta ^2}{2}-\zeta  \lambda +\frac{3 x^4}{4}-\frac{\lambda  x^3}{3}-\frac{3 x^2}{2}+\lambda  x\right)\right].
\end{equation}
As one can immediately see the integrals \eqref{eq:newpx} and \eqref{eq:vx} have no closed-form solution. One must therefore use approximation techniques to evaluate both $p(x)$ and $v(x)$. 
We performed a detailed numerical integration of Eq. \eqref{eq:vx}, using Mathematica's `NIntegrate()` function by sampling the state space one million times. There may be still the question of how one should choose the parameter $\lambda$ as defined in Eq. \eqref{eq:lambda} for the purpose of such a numerical simulation. As we mentioned in the introduction, in this paper, we are concerned with power-law inflation models such as large-field models \cite{elliscosmo} which have chaotic inflation as a key example. In the inflationary scenario, we have that $V(\phi) \ll 1$, which implies that $\phi \ll 1$. In this case, evaluating Eq. \eqref{eq:lambda} for $V = V_{n} \phi^{n}$, one obtains that
\begin{equation}
\lambda = \frac{\sqrt{3} n}{\sqrt{2} \phi}.
\end{equation}
Therefore, if $\phi \ll 1$, clearly $\lambda \gg 1$. Based on this, in our numerical simulations, we took many samples of $\lambda$ over the space $\lambda \in \left[10^2, 10^6\right]$. We display the results of our numerical integration in Figs. \ref{fig:fig1} and \ref{fig:fig2} below.
\begin{figure}[h]
\begin{center}
\includegraphics{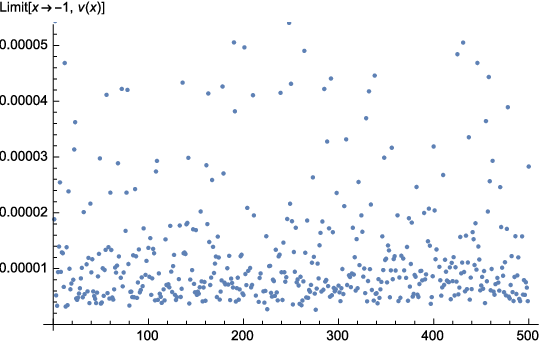}
\caption{Some results of the numerical integration of Eq. \eqref{eq:vx}, which clearly shows that $\lim_{x \to -1} v(x) < \infty$.}
\label{fig:fig1}
\end{center}
\end{figure}
\begin{figure}[h]
\begin{center}
\includegraphics{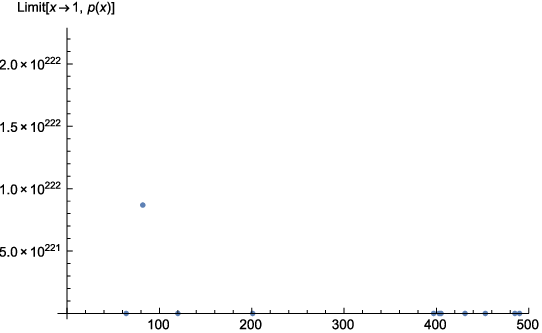}
\caption{Some results of the numerical integration of Eq. \eqref{eq:newpx}, which clearly shows that $\lim_{x \to 1} p(x) = \infty$. Note also that the way Mathematica apparently handles $\infty$ in its numerical integration algorithm is to provide unbelievably large numbers, in this case, on the order of $10^{222}$, which for all purposes is to be taken as $\infty$ in these computations.}
\label{fig:fig2}
\end{center}
\end{figure}
It is therefore quite clear based on our extensive numerical integration work, the results are that of condition \eqref{eq:cond3} above. Therefore, by Feller's explosion test, namely condition \eqref{eq:cond3}, solutions to Eq. \eqref{eq:Yev2} corresponding to an arbitrary power-law potential, explode in a finite time in the state space $Y \in \left[-1,1\right]$ with unitary probability.

We now analyze the solutions to the evolution equation for $X$, Eq. \eqref{eq:Xev1},
\begin{equation*}
X' = X \left(3 - 3X^2 + \sqrt{1 - X^2}\lambda\right),
\end{equation*}
where as mentioned above $\lambda \in \mathbb{R}$. We use a similar technique as above to determine the state space for this evolution equation. Namely, we obtain the condition that
\begin{equation}
1 - 2X^2 \leq 1 \Rightarrow X \in \mathbb{R}.
\end{equation}
Now, it is clear that Eq. \eqref{eq:Xev1} is an ordinary differential equation that contains no stochastic terms. We recall the following theorem from ordinary differential equations theory \cite{boyce}. Let the functions $f$ and $\partial f / \partial x$ be continuous in some rectangle $\alpha < \tau <\beta, \gamma < x < \delta$ containing the point $(\tau_0, x_0)$. Then, in some interval $\tau_{0} - h < \tau < \tau_{0} + h$ contained in $\alpha < \tau < \beta$, there exists a unique solution $x = \phi(\tau)$ of the initial value problem
\begin{equation}
\label{eq:ivpdef}
x' = f(\tau,x), \quad x(\tau_{0}) = x_{0}.
\end{equation}
As noted in \cite{boyce}, it is enough to guarantee existence, but not uniqueness of solutions on the basis of continuity of $f$ alone. Let us denote the right-hand-side of Eq. \eqref{eq:Xev1} by $f(X)$, such that
\begin{equation}
\label{eq:fX}
f(X) = 3X - 3X^3 + X \sqrt{1-X^2} \lambda,
\end{equation}
where $\lambda \in \mathbb{R}$. This implies that
\begin{equation}
\label{eq:fpX}
f'(X) =  X^2 \left(-9 -\frac{\lambda }{\sqrt{1-X^2}}\right)+\lambda  \sqrt{1-X^2}+3.
\end{equation}
Clearly, $f(X)$ is only valid for $-1 \leq X \leq 1$, while $f'(X)$ has singularities at $X = -1$ and $X = 1$. Therefore, by the above theorem, one can only guarantee existence and uniqueness of solutions in the open interval $X \in \left(-1,1\right)$ up to a finite explosion time, which is a much smaller subset of the physical state space of $X$ is $X \in \mathbb{R}$ by the inflation conditions we stated previously. 

\section{Conclusions}
In this paper, we considered the problem of existence and uniqueness of solutions to the Einstein field equations for a spatially flat FLRW universe in the context of stochastic eternal inflation where the stochastic mechanism is modelled by adding a stochastic forcing term representing Gaussian white noise to the Klein-Gordon equation. We showed that under these considerations, the Klein-Gordon equation actually becomes a stochastic differential equation. We further demonstrated that the existence and uniqueness of solutions to Einsteins equations depended on whether the coefficients of this stochastic differential equation obeyed global Lipschitz continuity and growth conditions.  We showed that for any choice of $V(\phi)$, the Einstein field equations were not well-posed, hence, any solution found to these equations was not guaranteed to be unique. Instead, we showed that the coefficients were at best, locally Lipschitz continuous in the physical state space of the Einstein equations, which only existed up to a finite explosion time. We then performed Feller's explosion test for an arbitrary power-law inflaton potential and proved that all solutions to the EFE explode in a finite time with probability one. This implies that the mechanism of stochastic inflation thus considered cannot be described to be eternal, since the very concept of eternal inflation implies that the process continues indefinitely. As mentioned in the introduction of this paper, this was the same conclusion reached in \cite{grattonturok}, namely that eternal inflation described by such a stochastic force term does not lead to future eternal behaviour. Further, we believe we have extended the results of \cite{grattonturok} to a more general case as the study of Gratton and Turok only considered a potential of the form $\lambda \phi^4$. Our work explicitly demonstrates that stochastic eternal inflation of the specific form considered here and in \cite{grattonturok} for \emph{any} power-law scalar field potential is not future eternal.

We therefore conclude that stochastic inflation based on a stochastic forcing term would not produce an infinite number of universes in some multiverse ensemble. In general, since the Einstein field equations in both situations were not well-posed, we conclude that the existence of a multiverse via the stochastic eternal inflation mechanism considered in this paper is still very much an open question that will require much deeper investigation.

\section{Acknowledgements}
The authors would like to thank the anonymous referees for their very helpful input and suggestions in revising the paper. In addition, ISK would like to thank George F.R. Ellis for his helpful comments at an early stage of the research of this topic, and for an initial reading of the manuscript.

\newpage
\bibliographystyle{alpha} 
\bibliography{sources}

\end{document}